\documentstyle[aps,prl,preprint,epsfig,12pt]{revtex}

\begin{document}

\epsfysize3cm
\epsfbox{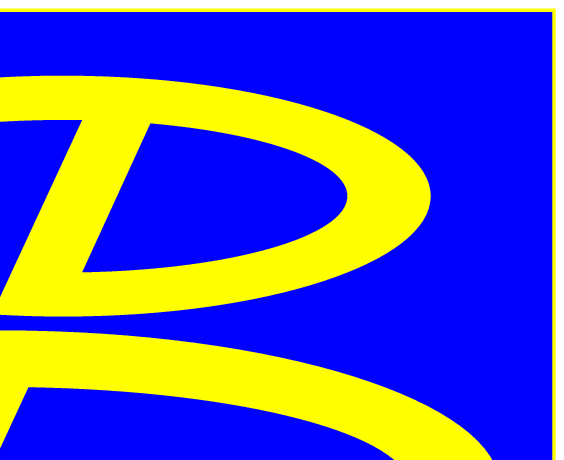}
\vskip -3cm
\noindent
\hspace*{12cm}KEK Preprint 2001-19 \\
\hspace*{12cm}Belle Preprint 2001-4 \\

\begin{center}

{\Large\bf
Observation of Cabibbo suppressed \\ $B \to D^{(*)}K^-$ decays at
Belle
\footnote{submitted to PRL}
}

\vskip 0.5cm

\tighten

%
%
\normalsize

The Belle Collaboration

\vskip 0.5cm

K.~Abe$^{10}$, 
K.~Abe$^{37}$, 
I.~Adachi$^{10}$, 
Byoung~Sup~Ahn$^{15}$, 
H.~Aihara$^{38}$, 
M.~Akatsu$^{20}$, 
G.~Alimonti$^{9}$, 
Y.~Asano$^{43}$, 
T.~Aso$^{42}$, 
V.~Aulchenko$^{2}$
% ...
\end{center}
\smallskip
\tighten
\begin{center}
% ...
$^{2}${Budker Institute of Nuclear Physics, Novosibirsk}\\
% ...
$^{9}${University of Hawaii, Honolulu HI}\\
$^{10}${High Energy Accelerator Research Organization (KEK), Tsukuba}\\
% ...
$^{15}${Korea University, Seoul}\\
% ...
$^{20}${Nagoya University, Nagoya}\\
% ...
$^{37}${Tohoku University, Sendai}\\
$^{38}${University of Tokyo, Tokyo}\\
% ...
$^{42}${Toyama National College of Maritime Technology, Toyama}\\
$^{43}${University of Tsukuba, Tsukuba}\\
% ...
\end{center}
\normalsize



\bigskip

 
%


\bigskip

\begin{abstract}
We report observations of the
Cabibbo-suppressed decays $B \to D^{(*)} K^-$
using a  10.4~fb$^{-1}$ data sample
accumulated at the $\Upsilon(4S)$ resonance
with the Belle detector at the KEKB $e^+ e^-$ storage ring.
The high-momentum particle
identification system of Belle is used  to
isolate  signals for
$B\to D^0 K^-$, $D^+K^-$, $D^{*0}K^-$ and $D^{*+}K^-$
from the $B\to D^{(*)}\pi^-$ decay processes
which have much larger branching fractions.
We report ratios of Cabibbo-suppressed to
Cabibbo-favored branching fractions of:
\smallskip
\renewcommand{\labelitemi}{ }
\begin{itemize}

\item{ }
\center{
${\cal B}(B^- \to D^0 K^-)$/${\cal B}(B^- \to D^0\pi^-)
= 0.079\pm0.009\pm0.006$;
}

\item{ }
\center{
${\cal B}(\bar{B^0} \to D^+ K^-)$/${\cal B}(\bar{B^0} \to D^+\pi^-)
= 0.068\pm0.015\pm0.007$;
}

\item{ }
\center{
${\cal B}(B^-\to D^{*0}K^-)/{\cal B}(B^-\to D^{*0}\pi^-)
= 0.078 \pm 0.019 \pm 0.009$;  {\rm and}
}

\item{ }
\center{
${\cal B}(\bar{B}^0\to D^{*+}K^-)/{\cal B}(\bar{B}^0\to D^{*+}\pi^-)=
0.074 \pm 0.015 \pm 0.006$.
} \\

\end{itemize}
\bigskip
\noindent
The first error is statistical and the second is systematic.  
These are the first reported observations of 
the $B\to D^+K^-$, $D^{*0}K^-$ and $D^{*+}K^-$ decay processes.
\end{abstract}
%

%
\bigskip
\leftline{~~~~~~~~~~~~~~~~PACS numbers: 12.15.Hh, 13.25.Hw}

{\renewcommand{\thefootnote}{\fnsymbol{footnote}}


\setcounter{footnote}{0}


\newpage



\narrowtext


Comprehensive tests of the Standard Model mechanism for $CP$ violation 
will 
ultimately
require measurements of the  $\phi_3$ angle of the Kobayashi-Maskawa 
unitarity triangle \cite{ref:KM}.  For this, many authors have
proposed the measurement of direct $CP$-violating asymmetries due to 
the interference between $b\to c$  and $b\to u$ transition
amplitudes in the Cabibbo suppressed  $B^-\to D^0K^-$ channel
as a theoretically clean way
to determine $\phi_3$~\cite{ref:GAMMA}.  As a first step
in this program, it is necessary to establish that the
Cabibbo suppressed decay modes exist and occur at the expected rates.

In the tree-level approximation, the branching fractions for
the Cabibbo suppressed decay processes $B\to D^{(*)}K^-$ are 
related to those for their $B\to D^{(*)}\pi^-$ counterparts 
\cite{ref:Expect} by
\begin{equation}
R\equiv \frac{{\cal B}(B\to D^{(*)}K^-)}{{\cal B}(B\to D^{(*)}\pi^-)}\simeq
\tan^2\theta_C (f_K/f_\pi)^2\simeq 0.074.
\end{equation}
Here $\theta_C$ is the Cabibbo angle, and $f_{K}$ and $f_{\pi}$ are the
meson decay constants. 
The only Cabibbo suppressed $B\to DK$ decay observed to date is
$B^-\to D^0 K^-$, which is reported  by the CLEO group to have
a ratio of branching fractions
$R = {{{\cal B}(B^-\to D^0K^-)}/{{\cal B}(B^-\to D^0\pi^-)}}
= 0.055\pm0.014\pm0.005$~\cite{ref:CLEO}, in agreement with expectations.

In this paper, we report the first observations of the Cabibbo suppressed
processes $B^-\to D^{*0}K^-$, $\bar{B}^0\to D^{*+}K^-$ and
$\bar{B}^0\to D^{+}K^-$, and a new measurement
of $B^-\to D^0 K^-$, using data collected at
the  $\Upsilon (4S)$ resonance with the Belle detector \cite{ref:belle}
at the KEKB asymmetric energy $e^+e^-$
collider \cite{ref:KEKB}.
The good high momentum particle identification capability of
the Belle detector enables us to extract  signals for
$B \to D^{(*)}K^-$ that are well separated from the more
abundant, Cabibbo favored $B \to D^{(*)}\pi^-$ processes.
The results are based on a 10.4~fb$^{-1}$ data sample that
contains 11.1 million $B\bar{B}$ pairs.

Belle is a general-purpose detector which includes
a 1.5~T superconducting solenoid 
magnet that surrounds the KEKB beam crossing point. 
Charged particle tracking covering approximately 90\%
of the total center of mass (cm) solid angle 
is provided by  a Silicon Vertex Detector (SVD),
consisting of three nearly cylindrical layers
of double-sided silicon strip detectors,
and a 50-layer Central Drift Chamber (CDC).
Particle identification is accomplished by combining
the responses from a Silica Aerogel \v{C}erenkov Counter 
(ACC)
and a Time of Flight Counter system (TOF)
with specific ionization ($dE/dx$) measurements in the CDC.
The combined response of the three systems provides 
$K/\pi$ separation of
at least `$2.5\sigma$ equivalent'  
for laboratory momenta up to 3.5 GeV/$c$.
For distinguishing the {\em prompt} kaons and pions
in $B\to D^{(*)}h^-$ ($h^- = K^-$ or $\pi^-$) decays, only
the ACC and $dE/dx$ are used since the
TOF system provides no significant kaon and pion separation 
at momenta relevant to this analysis. 
A CsI Electromagnetic Calorimeter (ECL) located inside
the solenoid coil
is used for $\gamma/\pi^0$ detection.

The $B \to D^{(*)}K^-$ 
processes have kinematic properties nearly identical  
to those of $B \to D^{(*)}\pi^-$ decays.
While the latter processes produce the
most significant backgrounds, they also 
provide control samples that
we use to establish cuts on kinematic variables, determine
the experimental resolutions, evaluate the
systematic errors, 
and normalize the results.

In this analysis, we require,
except for the $K_S \to \pi^+\pi^-$ decay daughters, 
that the charged tracks have a point 
of closest approach
to the interaction point within 
$\pm$5~mm in the direction perpendicular and $\pm$5~cm 
in the direction parallel to the beam axis.
For each charged track, the particle identification system is used to
determine a $K/\pi$ likelihood ratio $P(K/\pi)$
that ranges between 0 (likely to be $\pi$) and 1 (likely to be $K$).
We form candidate $D^0$ mesons using the
$K^-\pi^+$, $K^-\pi^+\pi^0$ and $K^-\pi^+\pi^+\pi^-$ decay modes, 
and $D^+$ mesons using $K^-\pi^+\pi^+$, $K_S\pi^+$, $K_S\pi^+\pi^+\pi^-$
and $K^-K^+\pi^+$ decays. 
(The inclusion
of charge conjugate states is implied throughout this report.)
For the assignment of charged kaons from $D$ decays, we
require $P(K/\pi)>0.3$ in reconstructing $D^0\to K^-\pi^+$  and
all other $D^0$ decay modes associated with $D^{*+}\to D^0\pi^+$,
otherwise we require $P(K/\pi)>0.7$.
Candidate $\pi^0$ mesons are reconstructed from
pairs of $\gamma$'s, each
with energy greater than 30~MeV, that have an invariant mass in the
range of 
$\pm 2 \sigma$ ($\sigma = 5.3$ MeV/$c^2$) of the measured $\pi^0$ mass
value 
and magnitude of the total three-momentum greater than 200 MeV/$c$.
For the slow $\pi^0$ from the $D^*\to D\pi^0$ decay
we only require that the invariant mass is in the range between 
$-14.4$ and $+10.8$ MeV of the $\pi^0$ mass.
The $K_S\to \pi^+\pi^-$ candidates are reconstructed from two 
oppositely charged tracks that form an invariant mass within 
$\pm 3\sigma$ ($\sigma=4.6$ MeV/$c^2$) of the measured $K_S$ mass value 
with a vertex which is displaced from the interaction point in the 
direction of the $K_S$ momentum.
The selected $\pi^0$ and $K_S$ candidates are kinematically constrained
to the nominal mass values.

For each $D$-meson decay topology,
we select particle combinations that have a reconstructed invariant mass
within $\pm 2.5\sigma_{D}$  
of the measured $D$ mass value, where $\sigma_{D}$ is
the $D$ mass resolution, which
varies from 5 to 13 MeV/$c^2$ depending on the decay mode.   
After selection, the $D$ candidates are subjected to a 
mass constrained kinematic fit.
For all modes except for $D^+\to K_S\pi^+$,
we further reduce backgrounds by a selection
on the quality of the vertex fit.

For $D^{*0}$ and $D^{*+}$ candidates, we use the
$D^{*0}\to D^0\pi^0$, and $D^{*+}\to D^0\pi^+$ and $D^+\pi^0$
decay channels.  We select events where the
mass difference between the $D\pi$ and $D$ particle combinations
is within $\pm 3\sigma$ of the expected value for 
$D^{*+}\to D^0\pi^+$, and $\pm 2\sigma$ for $D^{*0}\to D^0\pi^0$
and $D^{*+}\to D^+\pi^0$.  
A kinematic fit that constrains the $D^{*}$ mass to its nominal value
is applied to the events that satisfy the selection criteria.

When we isolate $B \to D^{(*)} h^-$ candidates, 
we use a quantity we call the {\em lab constrained mass}, 
$M_{lc}$, 
which is the $D^{(*)}h^-$
invariant mass calculated with the assumption
of two equal-mass particles from laboratory momenta:
$M_{lc} = \sqrt{(E^{lab}_{B})^{2} - (p_B)^{2}}$, where $p_B$ 
is the $B$ candidate's laboratory momentum and 
$E^{lab}_{B} = {1\over E_{ee}}(s/2 + \mbox{{\bf P}$_{ee}$} \cdot 
\mbox{{\bf P}$_B$} )$, where $s$ is square of the center of mass (cm)
energy, {{\bf P}$_B$} is the laboratory momentum vector of the 
$B$ meson candidate, 
and {\bf P$_{ee}$} and $E_{ee}$ are the laboratory  
momentum and energy of the $e^+e^-$ system, respectively.
To identify $B \to D^{(*)}K^-/D^{(*)}\pi^-$ samples 
we use the cm {\em energy difference}, which is defined as  
$\Delta E = E_{D^{(*)}}^{cm} + E_{\pi^-}^{cm} - E_{beam}^{cm}$,
where $E_{D^{(*)}}^{cm}$ is the cm energy of the $D^{(*)}$ candidate
and $E_{\pi^-}^{cm}$ is the cm energy of the prompt $h^-$
track calculated {\it assuming the pion mass}
and $E_{beam}^{cm}$ is the cm beam energy.
With this pion mass assumption, $B \to D^{(*)}\pi^-$
events peak at $\Delta E =0$, while the $D^{(*)} K^-$
peak is shifted to $\Delta E = -49$ MeV.
Typical $\Delta E$ resolutions obtained for $B\to D^{(*)}\pi^-$ and 
$B\to D^{(*)}K^-$ 
are 16 MeV and 19 MeV, respectively, hence we can distinguish 
these two processess by the $\Delta E$ distribution.  
For further analysis we accept $B$ candidates
with $5.27 < M_{lc} < 5.29$ GeV/$c^2$ 
and $-0.2 < \Delta E < 0.2$ GeV.

In the case of multiple entries from one event,
we choose the one with the smallest value of a
$\chi^2 $ that is determined from the differences between
measured and nominal values of $M_D$, $M_{lc}$ and, when appropriate,
$M_{D^*}-M_D$ and $M_{\pi^0}$.
For the latter, we
only consider the $\pi^0$ from
$D^{*0}\to D^0\pi^0$ and $D^{*+}\to D^+\pi^0$ decays.
Background events from
$e^+e^-\to q\bar{q}$ continuum processes are reduced 
using the normalized second Fox-Wolfram moment\cite{ref:FoxWolf}, 
$R_2$, and
the angle between the sphericity
axis of the $B$ candidate and the sphericity axis of the
rest of the particles in the event, $\theta_{sph}$.
For the events with $D^0\to K^-\pi^+$ decays and
all modes with $D^{*+}\to D^0\pi^+$, 
continuum backgrounds are small and  we only require $R_2<0.5$. 
This cut retains $96\%$ of the signal and rejects 
$25\%$ of the continuum.
For all other modes, we impose the additional requirement of
$|\cos\theta_{sph}|<0.75$, which retains $75\%$ of 
the signal and rejects $80\%$ of the continuum events.

We extract $B \to D^{(*)}K^-$ enriched samples by
applying a stringent particle
identification condition on the
prompt $h^-$, namely $P(K/\pi)>0.8$;
$B \to D^{(*)}\pi^-$ samples are selected  
with a criterion $P(K/\pi)<0.8$. 
The $\Delta E$ distributions for the $B\to D\pi^-$ 
[$B\to D K^-$ enriched] samples
are shown in  Figs.~\ref{fig:DeltaE_D}(a) and (b) [(c) and (d)].
The corresponding distributions for the $D^*$ channels
are shown in  Figs.~\ref{fig:DeltaE_Dstr}(a) through  (d).

Kaon and pion identification efficiencies $\varepsilon(K)$ and
$\varepsilon(\pi)$ are experimentally determined
from a kinematically selected sample of high momentum
$D^{*+}\to D^0\pi^+$, $D^0\to K^-\pi^+$ decays where the
$K^-$ and $\pi^+$ mesons from $D^0$ candidates are in the same
momentum and angular region as the prompt $h^-$ particle in
$B\to D^{(*)}h^-$ decay ($2.1 < p^{cm} < 2.5$ GeV/$c$).
With our $P(K/\pi)$ cut value of 0.8, 
the efficiencies are $\varepsilon(K) = 0.765\pm 0.006$ and
$\varepsilon(\pi) = 0.980\pm0.003$,
and the $\pi\to K$ fake rate is $2.0\pm 0.3\%$.

In Figs.~\ref{fig:DeltaE_D}(c) and (d), and
\ref{fig:DeltaE_Dstr}(c) and (d),
peaks around $\Delta E = -49$~MeV
correspond to $B\to D^{(*)}K^-$
decays while peaks around $\Delta E=0$ are due to feed-across
from misidentified $D^{(*)}\pi^-$ decays.
The area of the feed-across peak is
2.0\% of the $D^{(*)}\pi^-$ signal yield 
(in Figs.~\ref{fig:DeltaE_D}(a) and (b), and
\ref{fig:DeltaE_Dstr}(a) and (b)), which is
consistent with the $\pi\to K$ fake rate.

We determine the numbers of $D^{(*)} \pi^-$ events and 
the line shape parameters by fitting the $\Delta E$ distributions
for the $D^{(*)}\pi^-$ samples of 
Figs.~\ref{fig:DeltaE_D}(a) and (b),
and \ref{fig:DeltaE_Dstr}(a) and (b).
We represent the signal distributions using 
two Gaussian functions with different central values and widths. 
The background has two components.
One is due to contributions from $D\rho^-$, $D^*\rho^-$, 
and other $B$-meson decay modes~\cite{ref:Background}, 
which produce the complicated structures mostly seen 
at negative value of $\Delta E$,
and the other is due to continuum events, which populate the entire 
$\Delta E$ region.
The shapes of the $B\bar{B}$ backgrounds
are determined using Monte Carlo (MC) simulated events~\cite{ref:QQ};
those for the continuum backgrounds
are taken from the $\Delta E$ distributions
for events in the $M_{lc}$ side band regions 
($5.20 < M_{lc} < 5.26$ GeV/$c^2$).   
In the fit to the $D^{(*)}\pi^-$ 
sample $\Delta E$ distributions, 
we allow the peak positions, widths and normalization of the
signal functions to vary, as well as the normalization of
both the $B\bar{B}$ and continuum background contributions.
The fit results are shown in 
Figs.~\ref{fig:DeltaE_D}(a) and (b), and 
\ref{fig:DeltaE_Dstr}(a) and (b)
as  solid curves.  
The resulting numbers of $D^{(*)}\pi^-$ events
are given in Table~\ref{Table_Fit_Results}.

In the fits to the $D^{(*)}K^-$ enriched $\Delta E$ distributions,
the parameters of the two Gaussians for the $D^{(*)}K^-$ signal
are fixed at the values 
determined from fits to the $D^{(*)}\pi^-$ samples
with a kaon mass hypothesis applied to the prompt pion.
The relative position of the signal with respect to the original 
$D^{(*)}\pi^-$ signal is reversed. 
This procedure accounts for the kinematic shifts and smearing
of the $\Delta E$ peaks caused by the incorrect mass assignment.
For the feed-across from the $D^{(*)}\pi^-$ peak, we fix the
parameters at the values determined from the $D^{(*)}\pi^-$ fits.
In these fits,
the areas of the $D^{(*)} K^-$ signal and functions for the 
$D^{(*)}\pi^-$ feed-across
and the scaling factors of the continuum background shapes
are allowed to vary.
The $B\bar{B}$ background in the $D^{(*)}K^-$ enriched sample
has two components: modes which also contribute to the $D^{(*)}\pi^-$ 
sample with one track misidentified as a high momentum kaon and other 
Cabibbo suppressed modes.
The normalization of the feed-across from the first component 
is fixed to the fit result for the $B\bar{B}$ background
in the $D^{(*)}\pi^-$ sample multiplied by the measured pion fake rate.
The contributions from other $D^{(*)}K^{(*)}$ modes are determined from a 
Monte-Carlo simulation assuming that the branching 
fractions of the suppressed modes relative to the corresponding
$D^{(*)}\pi^-$/$D^{(*)}\rho^-$ modes are reduced by the usual Cabibbo factor.
The fit results are shown as solid curves in
Figs.~\ref{fig:DeltaE_D}(c) and (d), and 
\ref{fig:DeltaE_Dstr}(c) and (d).
The numbers of events in the $D^{(*)}K^-$ signal
and $D^{(*)}\pi^-$ feed-across peaks are listed in
Table~\ref{Table_Fit_Results}.  
Also listed in Table~\ref{Table_Fit_Results} are the statistical 
significance values for the $D^{(*)}K^-$ signals, 
which are determined from the likelihood
values of fits made to the $D^{(*)}K^-$ enriched $\Delta E$
distributions with the signal yield fixed to zero. 
For each channel, the statistical significance of the signal 
corresponds to at least five standard deviations.

Experimentally, the ratio of branching fractions
is given by
$$ 
R = {{N(B\to D^{(*)}K^-)}\over{N(B\to D^{(*)}\pi^-)}} \times
    {\eta(D^{(*)}\pi^-)\over\eta(D^{(*)}K^-)} \times
    {\varepsilon(\pi) \over \varepsilon(K)}
$$
\noindent
where $N$ and $\eta$ denote the numbers of events
and detection efficiencies for the indicated processes, and
$\varepsilon$'s are the prompt pion and kaon identification
efficiencies, respectively.
The signal detection efficiencies are determined using MC, 
and $\eta(D^{(*)}K^-)$ are approximately 5\% lower than 
$\eta(D^{(*)}\pi^-)$ due to the decay-in-flight effect of kaons.
Particle identification efficiencies are 
determined as described before.

Since the kinematics of the $B \to D^{(*)} K^-$ and
$B \to D^{(*)} \pi^-$ processes
are quite similar, most of the systematic effects
cancel in the ratios of branching fractions.
The main sources of systematic error that do not cancel
are the uncertainties in $K/\pi$ identification efficiencies
and the shape of the background in the $\Delta E$ distributions.
Using MC simulations,
we estimate the systematic error of the $K$ identification
due to differences in the angle-momentum
correlations of the $D^{*+}$ and signal samples
to be $2\%$.
To estimate the systematic error due to the parameterization of
the $\Delta E$ distributions, we use several different
methods of fitting. 
These include using linear background functions, 
fixing the $B\bar{B}$ backgrounds to MC calculated values, 
and forcing the parameters of $D^{(*)}K^-$ 
and/or $D^{(*)}\pi^-$ signals to deviate from their optimal values 
by $\pm 1\sigma$. 
The quadratic sums of those values are used as measures of 
the systematic errors from this source.
Compared to these, which range from 7.5 to 10.8$\%$,
other sources of systematic errors are
determined to be negligibly small.
The total systematic error for each channel
is taken to be the sum in quadrature
of the individual components.

The resulting  $R$ ratio measurements
are listed with their statistical and systematic errors
in Table~\ref{Table_Results}.
For the $B\to D^+K^-$, $D^{*0}K^-$ and $D^{*+}K^-$
decay processes, these are first measurements.
In all cases, the $R$ ratios are consistent
with the expected value given in Eq.~(1).

In summary,
by taking advantage of the good high momentum particle
identification capability of Belle,
we have observed signals for the Cabibbo suppressed decays
$B \to D^0K^-$, $D^+K^-$, $D^{*0}K^-$ and $D^{*+}K^-$
that are well separated from the more abundant
Cabibbo favored  $B \to D^{(*)}\pi^-$ processes.
We report values for the ratios
of branching fractions
$R={\cal B}(B\to D^{(*)} K^-)/{\cal B}(B\to D^{(*)}\pi^-)$
that agree, within errors, with 
naive model calculation. 

\medskip


We wish to thank the KEKB accelerator group for the excellent
operation of the accelerator.
We acknowledge support from
the Ministry of Education, Science, Sports and Culture of Japan
and the Japan Society for the Promotion of Science;
the Australian Research Council and the Australian Department of Industry,
Science and Resources;
the Department of Science and Technology of India;
the BK21 program of the Ministry of Education of Korea,
SRC program of the Korea Science and Engineering Foundation, 
and the Basic Science program of the Korea Research Foundation;
the Polish State Committee for Scientific Research
under contract No.2P03B 17017;
the Ministry of Science and Technology of Russian Federation;
the National Science Council and the Ministry of Education of Taiwan;
the Japan-Taiwan Cooperative Program of the Interchange Association;
and  the U.S. Department of Energy.


\newpage


\begin{table}[b]
   \caption{The fit results for the numbers of $D^{(*)}\pi^-$ and $D^{(*)}K^-$
 events, feed-across from $D^{(*)}\pi^-$ to $D^{(*)}K^-$ enriched 
 sample and the statistical significance of $D^{(*)}K^-$ signal.}
   \label{Table_Fit_Results}  
   \begin{center}
   \begin{tabular}{lllll}  
       & $D^{(*)}\pi^-$ & $D^{(*)}K^-$ & $D^{(*)}\pi^-$ & stat. \\ 
       & events         & events       & feed-across      & sig.  \\ \hline 
$B^-\to D^0h^-$    & $2402.8\pm97.8$ & $138.4\pm15.5$ & $52.0\pm11.4$ & 11.7 \\
$\bar{B^0}\to D^+h^-$ & $681.9\pm32.1$ & $33.7\pm7.3$ & $10.4\pm4.9$  & 6.1 \\
$B^-\to D^{*0}h^-$    & $584.8\pm32.4$ & $32.8\pm7.8$ & $6.8\pm4.9$   & 5.8 \\
$\bar{B^0}\to D^{*+}h^-$ & $640.9\pm30.8$ & $36.0\pm7.1$ & $21.0\pm5.7$ & 7.6 
   \end{tabular} 
  \end{center}
\end{table}

\begin{table}[b]
   \caption{The resulting branching fraction ratio measurements. 
    The first error is statistical and the second is systematic.
}
   \label{Table_Results}  
   \begin{center}
   \begin{tabular}{ll}  
$~~~{\cal B}(B^- \to D^0 K^-)$/${\cal B}(B^- \to D^0\pi^-) $ & 
$=~~0.079\pm0.009\pm0.006$~~~ \\
$~~~{\cal B}(\bar{B^0} \to D^+ K^-)$/${\cal B}(\bar{B^0} \to D^+\pi^-) $ & 
$=~~0.068\pm0.015\pm0.007$ ~~~\\
$~~~{\cal B}(B^-\to D^{*0}K^-)/{\cal B}(B^-\to D^{*0}\pi^-) $ &
$=~~0.078 \pm 0.019 \pm 0.009$~~~ \\
$~~~{\cal B}(\bar{B}^0\to D^{*+}K^-)/{\cal B}(\bar{B}^0\to D^{*+}\pi^-) $ &
$=~~0.074 \pm 0.015 \pm 0.006~~~$ 
   \end{tabular} 
  \end{center}
\end{table}



\begin{figure}[hb]
  \begin{center}
    \mbox{\psfig{figure=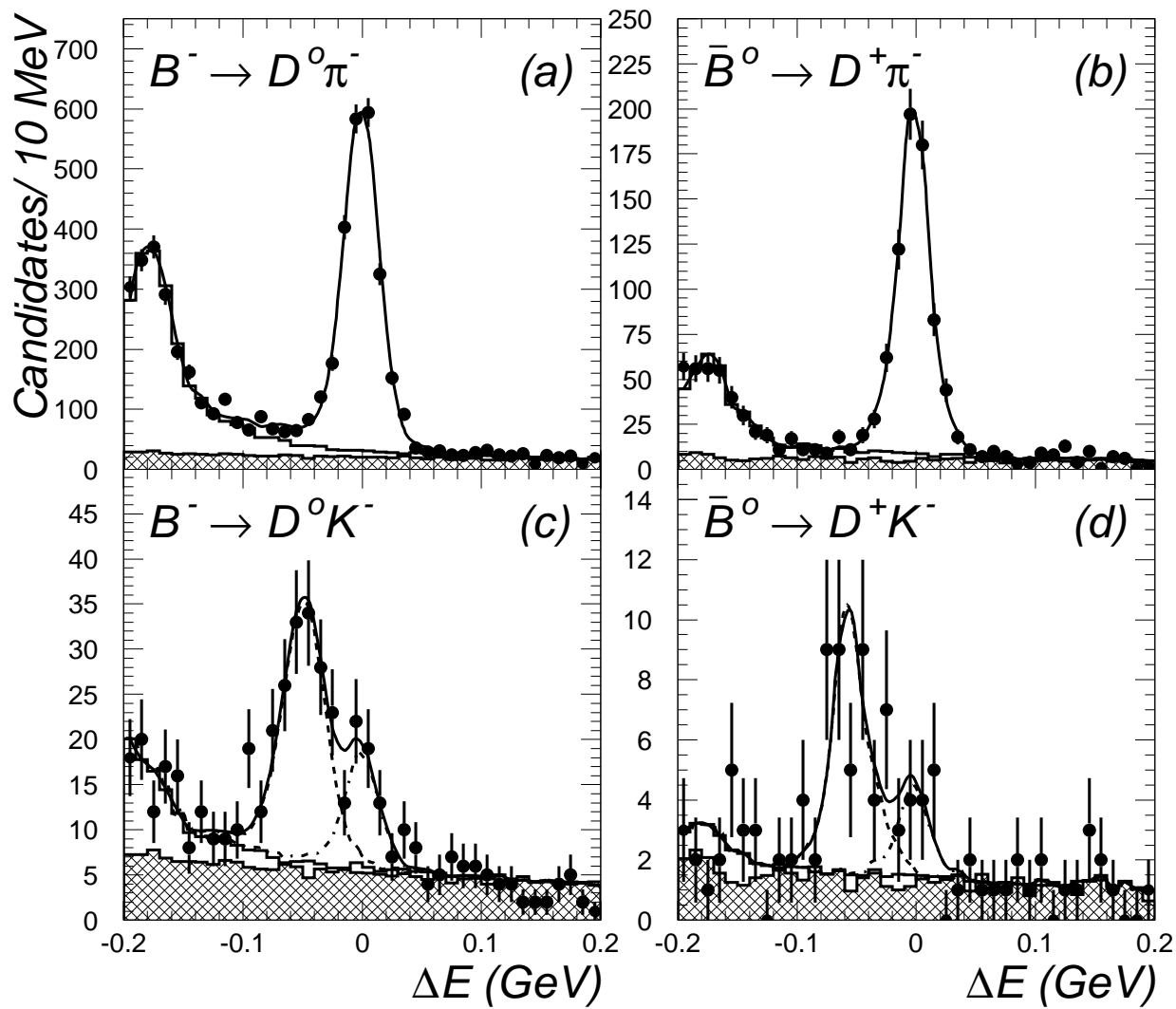,height=14cm}}
  \end{center}
  \caption{The $\Delta E$ distributions for the
          {\bf (a)} $B^-\to D^0\pi^-$ and
          {\bf (b)}$\bar{B}^0\to D^{+}\pi^-$ 
	samples, and the
          {\bf (c)} $B^-\to D^0 K^-$ and
          {\bf (d)}$\bar{B}^0\to D^{+}K^-$
        enriched samples, where in each case a pion mass is
       assigned to the $\pi^-/K^-$ track candidate.
      The points with error bars are the data, the curves
      show the results of fits that are described in the text.
     The open histograms are the sums of background
     functions  scaled to fit the data and the hatched histogram
     indicates the continuum component of the background.
}
  \label{fig:DeltaE_D}
\end{figure}

\begin{figure}
  \begin{center}
    \mbox{\psfig{figure=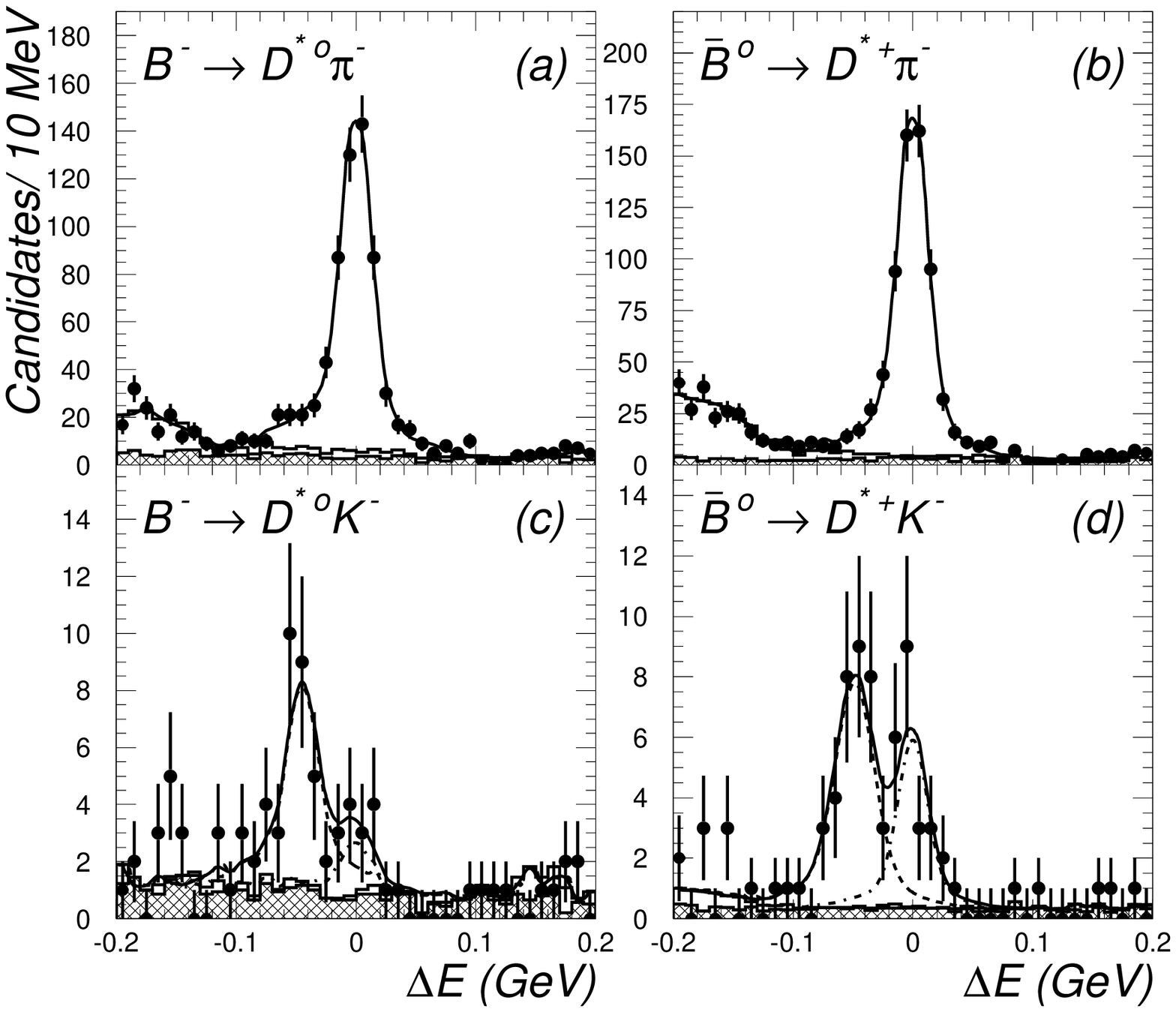,height=14cm}}
  \end{center}
  \caption{The $\Delta E$ distributions for the
          {\bf (a)} $B^-\to D^{*0}\pi^-$ and
          {\bf (b)}$\bar{B}^0\to D^{*+}\pi^-$ 
	samples, and the
          {\bf (c)} $B^-\to D^{*0} K^-$ and
          {\bf (d)}$\bar{B}^0\to D^{*+}K^-$
        enriched samples, where in each case a pion mass is
       assigned to the $\pi^-/K^-$ track candidate.
      The curves and histograms are the same as those in
      Fig.~\ref{fig:DeltaE_D}.
}
  \label{fig:DeltaE_Dstr}
\end{figure}


\bigskip\bigskip



\begin{thebibliography}{99}

\bibitem{ref:KM} 
M. Kobayashi and K. Maskawa, Prog. Theor. Phys. {\bf 49} 652 (1973).

\bibitem{ref:GAMMA} 
M.~Gronau and D.~Wyler, Phys. Lett. {\bf B265} 172 (1991);~
I.~Dunietz, Phys. Lett. {\bf B270} 75 (1991);~ 
D.~Atwood, I.~Dunietz and A.~Soni, Phys. Rev. Lett. {\bf 78} 3257 (1997).

\bibitem{ref:Expect} 
This relation assumes the validity of
factorization and flavor-$SU(3)$ symmetry.  
The numerical value is determined from adjusting the measured result
for the ratio $\tau^-\to K^-\nu$
and $\pi^-\nu$ decays corrected for the phase space effect.

\bibitem{ref:CLEO} 
M.Athanas {\it et al.} (CLEO collaboration), 
Phys. Rev. Lett. {\bf 80} 5493 (1998).

\bibitem{ref:belle} 
`The Belle Detector', Belle collaboration,
KEK Progress Report 2000--4, to be published in Nucl. Inst. Meth. 

\bibitem{ref:KEKB} 
KEKB B-Factory Design Report, KEK Report 95--7 (1995); 
Y. Funakoshi {\it et al.}, Proc. 2000 European Particle Accelerator 
Conference, Vienna (2000).








\bibitem{ref:FoxWolf}
G. Fox and S. Wolfram, Phys. Rev. Lett. {\bf 41} 1581 (1978).

\bibitem{ref:Background}
The $\Delta E$ plots for $D\pi^-$ and $D K^-$ have additional
background at negative $\Delta E$ values from
$B\to D^*\pi^-$  decays.

\bibitem{ref:QQ}
`QQ - The CLEO Event Generator', \\
http://www.lns.cornell.edu/public/CLEO/soft/QQ (unpublished).




\end{thebibliography}
\end{document}